\documentclass{sig-alternate}

\usepackage{subfigure}
\usepackage{algorithmic}
\usepackage{algorithm}
\usepackage{balance}
\usepackage{verbatim}

\begin{document}
\title{Bridge Bounding: A Local Approach for Efficient Community Discovery in Complex Networks}
%
% You need the command \numberofauthors to handle the "boxing"
% and alignment of the authors under the title, and to add
% a section for authors number 4 through n.
%
% Up to the first three authors are aligned under the title;
% use the \alignauthor commands below to handle those names
% and affiliations. Add names, affiliations, addresses for
% additional authors as the argument to \additionalauthors;
% these will be set for you without further effort on your
% part as the last section in the body of your article BEFORE
% References or any Appendices.

\numberofauthors{3}
%
% Put no more than the first THREE authors in the \author command

% NOTE: All authors should be on the first page. For instructions
% for more than 3 authors, see:
% http://www.acm.org/sigs/pubs/proceed/sigfaq.htm#a18

\author{
%% e-mail address with \email.
\alignauthor Symeon Papadopoulos\titlenote{This author is also affiliated with the Aristotle University of Thessaloniki.}\\
       \affaddr{Informatics and Telematics Institute}\\
       \affaddr{P.O.Box 60361, 57001, Thermi}\\
       \affaddr{Thessaloniki, Greece}\\
       \email{papadop@iti.gr}
\alignauthor Andre Skusa\\
       \affaddr{Lycos Europe GmbH}\\
       \affaddr{ Carl-Bertelsmann-Str. 29}\\
       \affaddr{33311 G\"{u}tersloh, Germany}\\
       \email{andre.skusa@lycos-europe.com}
\alignauthor Athena Vakali\\
       \affaddr{Department of Informatics}\\
       \affaddr{Aristotle University of Thessaloniki}\\
       \affaddr{54124 Thessaloniki, Greece}\\
       \email{avakali@csd.auth.gr}
\and \alignauthor Yiannis Kompatsiaris\\
       \affaddr{Informatics and Telematics Institute}\\
       \affaddr{P.O.Box 60361, 57001, Thermi}\\
       \affaddr{Thessaloniki, Greece}\\
    \email{ikom@iti.gr}
 \alignauthor Nadine Wagner\\
       \affaddr{Lycos Europe GmbH}\\
       \affaddr{ Carl-Bertelsmann-Str. 29}\\
       \affaddr{33311 G\"{u}tersloh, Germany}\\
    \email{nadine.wagner@lycos-europe.com}
}

\date{30 October 2008}

\maketitle
\begin{abstract}
The increasing importance of Web 2.0 applications during the last
years has created significant interest in tools for analyzing and
describing collective user activities and emerging phenomena within
the Web. Network structures have been widely employed in this
context for modeling users, web resources and relations between
them. However, the amount of data produced by modern web systems
results in networks that are of unprecedented size and complexity,
and are thus hard to interpret. To this end, \emph{community
detection} methods attempt to uncover natural groupings of web
objects by analyzing the topology of their containing network.

There are numerous techniques adopting a \emph{global} perspective
to the community detection problem, i.e. they operate on the
complete network structure, thus being computationally expensive and
hard to apply in a streaming manner. In order to add a \emph{local}
perspective to the study of the problem, we present \emph{Bridge
Bounding}, a local methodology for community detection, which
explores the local network topology around a seed node in order to
identify edges that act as boundaries to the local community. The
proposed method can be integrated in an efficient global community
detection scheme that compares favorably to the state of the art. As
a case study, we apply the method to explore the topic structure of
the \emph{LYCOS iQ} collaborative question/answering application by
detecting communities in the networks created from the collective
tagging activity of users.
\end{abstract}

% A category with only the three required fields
%A category including the fourth, optional field follows...



\section{Introduction}
Network structures (also called \emph{graphs} in mathematical
literature) and the associated analysis methods have long emerged as
a valuable tool for modeling and analyzing the relations among
objects in a variety of established scientific disciplines, e.g.
social sciences and biology \cite{www:newman03}. Recent years
however have witnessed a substantial adoption of network analysis
techniques in the field of computer science, and more specifically,
in the modeling and analysis of massive data sets produced by online
information systems, such as Web 2.0 applications.

In the field of network research, the problem of \emph{community
detection} has lately attracted significant interest since
identifying the community structure of large networks can improve
our understanding of the complex relations that exist among their
elements. The origins of this problem can be traced in the fields of
citation study \cite{www:garfield79}, bibliometrics
\cite{www:white89} and social network analysis \cite{www:scott00}.
More recently, this problem has been restated in the context of web
graphs, i.e. the networks created from mapping the web hyperlink
structure to the directed network model. Two seminal web community
definitions were formulated by Kumar et al.\ \cite{www:kumar99} and
Flake et al.\ \cite{www:flake00}. According to the first, a
community is a \emph{dense directed bipartite subgraph} of the web
graph \cite{www:kumar99}. The latter definition states that a
community is \emph{a vertex subset of the graph such that each of
its members has at least as many edges to other members of the
community as it does to non-member vertices} \cite{www:flake00}.
Although these two community definitions are different, they both
result in the formulation of community detection as a problem of
finding a partition of a graph into subgraphs that maximizes some
measure of within-subgraph density.

Due to the extremely high complexity of providing an exact solution
to the community detection problem for the complete
network\footnote{The problem is believed to be NP-complete
\cite{www:newman04a}.}, several attempts have been made to derive
approximate solutions at reduced computational costs, with some of
the most efficient techniques having a complexity of $O(nlog^2n)$
\cite{www:newman04b} and $O(m+n)$ \cite{www:wu03} for networks of
$n$ nodes and $m$ edges. Despite being very efficient, most of the
existing approaches adopt a \emph{global} perspective, i.e. they
operate on the full network, in order to output the detected
community structure. In practice, however, there is frequently a
need to explore the network structure at a \emph{local} level, e.g.
in interactive network visualization \cite{www:aris06} and
information retrieval applications \cite{www:song08}. Such
applications impose severe constraints on the response time of the
underlying network analysis processes, thus prohibiting the use of
global community detection methods. To date, only few methods have
been proposed that can be used for community detection at a local
level \cite{www:bagrow05, www:wu03}. However, they are either
unsuitable for networks of scale-free topology (frequently emerging
in practice) \cite{www:bagrow05} or are not local by design, thus
not achieving maximum efficiency when applied as local
\cite{www:wu03}.

The situation described above motivated us to introduce a
methodology for performing community detection at a local level; we
call the proposed methodology \emph{Bridge Bounding}. Bridge
Bounding initiates the community detection process from a seed node
in the network and progressively attaches neighboring nodes to the
community as long as the edges connecting these nodes do not act as
\emph{boundaries}. Thus, community detection is formulated as a
problem of identifying edges that act as community boundaries,
(which we also call \emph{bridges}, since they connect communities
of the network to each other \cite[p.140]{www:denooy05}). This
problem is tackled by means of \emph{local network topology
functions}, i.e. functions that examine the network structure around
an edge (local network topology) and produce a measure of the extent
that these edges act as bridges. An example of such a function is
the \emph{edge clustering coefficient} \cite{www:radicchi04}. In
that way, we ensure that the proposed approach has low complexity
and at the same time is capable of precisely identifying community
boundaries.

In order to demonstrate the benefits of our approach, we applied it
to both synthetic and real networks. As a first step, we validated
Bridge Bounding by testing its performance on the known community
structure of synthetic networks and comparing it with the widely
cited approach of Girvan and Newman \cite{www:girvan02}. The
proposed method could successfully detect the synthetic communities
across a variety of network generation parameters and achieved
equivalent or better performance than the competing method, while
being computationally much more efficient. Subsequently, we employed
Bridge Bounding to explore the community structure of two tag
networks, English and German, created by tags used to annotate
questions in the LYCOS iQ question/answering application\footnote{We
collected data from both the German (http://iq.lycos.de) and the
English (http://iq.lycos.co.uk) version of the application}. A set
of tag communities consisting of semantically related tags were
extracted, thus revealing the structure of topics associated with
the collective question-answering activity of users. The extracted
tag community structure can be exploited for improved topic interest
monitoring and automatic tag recommendation to users of the
application.

The rest of the paper is structured as follows. Section
\ref{sec:relatedWork} presents an overview of existing work in the
field of community detection in complex networks. Subsequently, the
formal description of the proposed community detection methodology
is presented in Section \ref{sec:methodology}. Section
\ref{sec:evaluation} presents the results and insights we obtained
by applying Bridge Bounding both to synthetic networks with known
community structure and to the LYCOS iQ tag network. Finally,
Section \ref{sec:conclusions} summarizes the basic contributions of
the paper and delineates our future work.

%%%%%%%%%%%%%% RELATED WORK %%%%%%%%%%%%%%%%%%%%%%
\section{Related Work}
\label{sec:relatedWork} The problem of community finding in large
complex networks has attracted considerable research interest for
some time now. Its origins can be traced back to the first studies
of the hyperlink structure of the web, e.g. to the observation of
Gibson et al.\ \cite{www:gibson98} that communities emerge
spontaneously around authoritative web pages which are identified by
means of hub pages. Then, the works by Kumar et al.\
\cite{www:kumar99} and Flake et al.\ \cite{www:flake00} formally
defined and systematically tackled the problem of community
detection. In the following, we provide a list of existing methods
for community detection classified according to the approach they
adopt. A more detailed discussion of existing community detection
methods is contained in the survey by Danon et al.\
\cite{www:danon05a}.

\textbf{Subgraph enumeration.} Kumar et al.\ \cite{www:kumar99}
consider communities as dense bipartite subgraphs of the web (seen
as a directed graph). A natural way to identify dense subgraph
structures is by means of graph partition enumeration. In order to
drastically reduce the vast number of subgraphs that are possible by
complete enumeration, the authors employ a series of heuristic
pruning techniques. An extension of this definition led to the
notion of $\gamma$-dense communities \cite{www:dourisboure07}, which
can be efficiently discovered based on more sophisticated subgraph
enumeration and pruning criteria.

\textbf{Maximum flow.} Flake et al.\ \cite{www:flake00} define
communities as subsets of vertices that have more links (undirected)
to each other than to the rest of the network nodes. To detect such
communities on the web, they integrate a maximum flow strategy with
an iterative crawling process. A stricter community definition was
considered by Ino et al.\ \cite{www:ino05} and a technique was
devised to detect them that was based on both the maximum flow
algorithm and an iterative graph partitioning and contraction
process.

\textbf{Divisive-Agglomerative methods.} According to Girvan and
Newman \cite{www:girvan02}, the community structure of a large
network should be revealed by progressively removing edges with high
edge betweenness, i.e. by following a divisive approach. Following
the same approach but with the use of different measures, namely the
edge \emph{clustering coefficient} and the \emph{bridging
centrality}, Radicchi et al.\ \cite{www:radicchi04} and Hwang et
al.\ \cite{www:hwang08}, respectively, could uncover the underlying
community structure of complex networks. Later, the measure of
\emph{modularity} was defined by Newman and Girvan, as a means to
quantify the quality of a network partition into communities
\cite{www:newman04a}. More specifically, modularity reflects the
extent to which a given network partition is characterized by higher
intra-community density in comparison to the one that would be
observed in a random partition of the same network. Building upon
this measure, the methods by Newman \cite{www:newman04b} and Clauset
et al.\ \cite{www:clauset04} describe efficient implementations of
community detection by means of agglomerative strategies.

\textbf{Seed-based Flooding.} An alternative approach to assigning
the nodes of a network to communities was presented by da Fontoura
Costa in \cite{www:costa04}. There, the community detection process
starts from a set of \emph{hub nodes} and is implemented as a
parallel flooding process emanating from the hubs. Although being
seed-based, the technique in \cite{www:costa04} is not local since
it requires simultaneous discovery of all communities in a network.
Thus, a local method for community finding was described by Bagrow
and Bollt \cite{www:bagrow05}. The authors consider an expanding
neighborhood around the starting node (which they call $l$-shell) to
constitute the community around it. In order to finish the expansion
process, the authors employ a criterion quantifying the change in
the \emph{total emerging degree} of the community
\cite{www:bagrow05}.

\textbf{Hybrid.} A combined strategy for community detection is
provided by Du et al.\ \cite{www:du07}. The authors consider a
three-step community detection process: (a) detection of maximal
cliques (subgraph enumeration), (b) initial network partition by
progressive expansion of the maximal cliques (flooding) and (c)
adjustment of the original partition in order to maximize
modularity.

Most of the methods presented above are global, meaning that they
need to process the whole network in order to output the identified
community structure. Even though some of these methods achieve low
complexity (linear to the size of the network), their use is still
prohibitive, when there is need for extremely responsive community
detection, e.g. in interactive exploration of large networks, which
can be only feasible by means of local processing of the network. We
could only find two local methods \cite{www:bagrow05, www:wu03} that
are suitable for identifying communities within such applications.
However, we consider the first of those \cite{www:bagrow05} as
unsuitable for graphs of scale-free nature (since the $l$-shell
would contain the whole graph after just few expansion steps), and
the second \cite{www:wu03} as not achieving maximum efficiency,
since it is not local by design (i.e. there are redundant
computation steps when applying the method locally). We consider
that our proposed methodology addresses the community detection
problem from a local perspective in a more intuitive and efficient
way.

Most existing community detection methods, to date, have been
applied to two types of networks: (a) networks created from crawling
part of the web and (b) networks reflecting the social relations
and/or interactions among people. Recently, there has also been some
work highlighting the value of community detection in tagging
systems.\footnote{Community detection is frequently termed
\emph{clustering} in the respective literature.} Part of the case
study in \cite{www:brooks06}, which mainly deals with the evaluation
of the effectiveness of tags as a means to annotate blog articles,
describes the induction of a tag hierarchy by means of a standard
hierarchical clustering scheme based on cosine similarity. In
another study \cite{www:song08}, a method based on Spectral
Recursive Embedding is proposed to carry out multi-clustering on the
two bipartite graphs formed by the documents-words and
documents-tags interrelationships in order to improve the precision
of tag recommendation. Finally, Cattuto et al.\ \cite{www:cattuto08}
exploit the tag overlap between online resources in order to
identify resource communities by means of spectral methods. In this
work, we apply our proposed methodology to the tag network created
from the collective tagging activity of the LYCOS iQ users. In that
way, we show that the topological properties of tag networks can be
exploited to extract tag groups that are semantically related to
each other.

\section{Methodology}
\label{sec:methodology} In this section, we will first (Section
\ref{sec:basic_notation}) present the basic notations and
definitions from graph theory that are necessary to formalize the
problem of community detection. Then, we will introduce the Bridge
Bounding community detection methodology in Section \ref{sec:bb}.

\subsection{Basic notation and definitions}
\label{sec:basic_notation} We consider undirected graphs $G =
(V,E)$, where $V$ is the set of vertices and $E \subseteq V \times
V$ is the set of edges connecting the vertices. An edge connecting
nodes $i, j \in V$ is denoted as $e_{ij}$.
\begin{comment}
Also, we assume that each edge $e_{st} \in E$ between two vertices
$s, t \in V$ is mapped to a weight $w_{st}$. For two unconnected
vertices $v_x, v_y \in V$, it is $w_{xy}=0$.
\end{comment}
For a vertex $s$ of the graph, we consider its neighborhood $N(s)$
consisting of all vertices which are directly connected to $s$, i.e.
$\forall n \in N(s): e_{sn} \in E$. We define the degree of vertex
$v$ as $d(v)=|N(v)|$. In a similar way, the neighborhood of an
\emph{edge} $e_{st}$ consists of all edges that share at least one
endpoint with $e_{st}$, $N(e_{st})=\{e_{xy}|\{x,y\} \bigcap \{s,t\}
\neq \oslash\}$.

Global community detection algorithms process a graph $G$ in order
to partition the graph into a set of communities, $\textbf{P} \equiv
\{C_0, C_1, ..., C_K\}$, where $C_i \subseteq V$. When the
communities produced by a method are mutually exclusive, then $C_i
\bigcap C_j = \oslash$, $\forall i, j \in \{1, 2, ..., K\}$, with $i
\neq j$. During the community detection process, we consider the set
of nodes $C_U \in \textbf{P}$ comprising all nodes that have not
been assigned to any community until that moment. For convenience,
we also employ the mapping $g_C: V \rightarrow \textbf{P}$, which
returns the community a vertex is assigned to (or $C_U$ if the
vertex has not been assigned to any community yet).

Local methods for community detection adopt a seed-based approach,
i.e. given $G$ and a node $s$ in the graph, a local method will
produce a community $C_s$ around the node. It is possible to induce
a global community detection method based on a local one by
repeatedly applying the local community detection method to randomly
selected nodes from $C_U$ until this set is empty (i.e. all nodes of
the graph have been assigned to some community). In the context of
our evaluation (Section \ref{sec:evaluation}), we are going to
induce such a global community detection scheme by employing the
local Bridge Bounding method, which we describe in Section
\ref{sec:bb}. We will refer to this scheme as \emph{progressive}
community detection.

\subsection{Community detection by Bridge Bounding}
\label{sec:bb} Bridge Bounding is based on a simple strategy in
order to identify the community $C_s$ surrounding a seed node $s$. A
formal description of this strategy is presented below, in Algorithm
\ref{alg:local_bb}. Starting from $s$, each node $n$ belonging to
the neighborhood of $s$ is considered a member of $C_s$ as long as
it meets two conditions (line \ref{alg:condition} of the algorithm):
(a) it is not already member of another community and (b) the edge
connecting it to $s$ is not a community boundary, i.e. not a
\emph{bridge} (in the sense of \cite[p.140]{www:denooy05}). Then,
all neighbors of the newly assigned nodes (the frontier set $F$) are
checked against the same conditions and are attached to $C_s$ (line
\ref{alg:pushToF}, lines \ref{alg:addToC}-\ref{alg:removeFromNonC})
if they meet them. This process is repeated until it is not possible
to attach additional nodes to $C_s$ (line \ref{alg:Fempty}). Thus,
Bridge Bounding is equivalent to a flooding process, similar to the
one described in \cite{www:costa04}, which stops when all nodes
belonging to its frontier are adjacent to a bridge (community
boundary).

\begin{algorithm}
\caption{LocalCommunityDetection} \label{alg:local_bb}
\begin{algorithmic}[1]
\REQUIRE Seed node $s \in G=(V,E)$ \REQUIRE Community mapping $g_C:
V \rightarrow \textbf{P}$ \REQUIRE Bridge function $b: E \rightarrow
[0.0, 1.0]$ \STATE $C_s = \oslash$ \STATE Frontier set $F=\{s\}$
\WHILE[$F$ is non-empty]{$|F|>0$} \label{alg:Fempty} \STATE $c
\leftarrow F$.pop() \STATE $C_s \leftarrow C_s \bigcup \{c\}$
\label{alg:addToC} \STATE $C_U \leftarrow C_U \backslash \{c\}$
\label{alg:removeFromNonC} \FORALL{$n \in N(c)$ such that
$e_{cn}=(c,n) \in E$ } \IF {$g_C(n)=C_U$ and $b(e_{cn}) \leq B_L$}
\label{alg:condition} \STATE $F$.push($n$) \label{alg:pushToF}
\ENDIF \ENDFOR \ENDWHILE \STATE $\textbf{P} \leftarrow \textbf{P}
\bigcup C_s$
\end{algorithmic}
\end{algorithm}

The quality of the community structure output by Bridge Bounding is
entwined with the success of quantifying the bridging behavior of
edges. Let us consider the function $b: E \rightarrow [0, 1]$, which
maps edges to real numbers in the given interval, to quantify the
extent to which they act as bridges. In order for Bridge Bounding to
make a binary decision on whether an edge $e$ is a bridge or not (in
order to stop or continue the community flooding process along this
edge), the output of the bridging function, $b(e)$, is compared
against some threshold $B_L$ (which can be derived by analysis of
the distribution of $b(e)$ values as will be shown later).

The problem of quantifying the bridging behavior of edges on a graph
has been already studied and several measures based on graph
topology have been developed with the goal of capturing the extent
to which an edge acts as a bridge between different communities. One
of the first attempts to define $b(e)$ was by means of its
\emph{betweenness centrality} as described in \cite{www:newman04a}.
For a given edge $e$, its betweenness centrality is defined as the
fraction of shortest paths running along the edge, $\sigma_{st}(e)$
to the number of all possible shortest paths $\sigma_{st}$ between
$s$ and $t$.

\begin{equation}
\label{eq:edge_betweeness} b_{\Phi}(e_{st})=\Phi(e_{st})= \sum_{s
\neq t \in V}{\frac{\sigma_{st}(e)}{\sigma_{st}}}
\end{equation}

An extension to this measure, called \emph{bridging centrality},
appeared in \cite{www:hwang08}. Bridging centrality was defined as
the rank product of the edge betweenness (Equation
\ref{eq:edge_betweeness}) and the edge \emph{bridging coefficient},
which made use of the local network topology to quantify the extent
to which an edge acts as a bridge.

\begin{comment}
\begin{equation}
\label{eq:bridging_coef} \Psi(e_{st})=\frac{d(s) \Psi(s)+d(t)
\Psi(t)}{(d(s)+d(t))(|N(s) \bigcap N(t)|+1)}
\end{equation}

The coefficient $\Psi(v)$ in Equation \ref{eq:bridging_coef} is the
bridging coefficient of a node:

\begin{equation}
\Psi(v)=\frac{1}{d(v)} \sum_{n \in N(v)} \frac{\delta(n)}{d(n)-1}
\end{equation} where $\delta(v)$ is the number of edges leaving the $N(v)$-induced
subgraph.
\end{comment}

The measures of betweenness and bridging centrality are global
bridging measures, i.e. they are computed by processing the whole
graph. To reduce the computational requirements, one may consider
local bridging measures, e.g. the \emph{edge-clustering coefficient}
\cite{www:radicchi04}: \begin{equation}
C_{st}^{(3)}=\frac{z_{st}^{(3)}}{min[(d(s)-1),(d(t)-1)]}
\end{equation} where $z_{st}^{(3)}$ is the number of triangles containing that
edge. Note that the larger the clustering coefficient is, the less
the edge acts like a bridge. Hence, we define the \emph{local
bridging} of an edge as:

\begin{equation}
b_{L}(e_{st})=1-C_{st}^{(3)}=1-\frac{|N(s) \bigcap
N(t)|}{min[(d(s)-1),(d(t)-1)]} \label{eq:local_bridging}
\end{equation}

In order for $b_L(e)$ to have a low value, the two endpoints of $e$
need to have a lot of common neighbors (relative to their degree).
Effectively, this means that in order to move from one of the
endpoints to the other, one has multiple options in addition to $e$.
Thus, $e$ is considered as an intra- (or within-) community edge. In
the opposite case, when the two endpoints of a bridge have very few
or no neighbors in common, then this edge is crucial for the
connection between its endpoints. For that reason, we consider in
the latter case, where $b_L(e)$ has a high value, that $e$ is an
inter-community edge or bridge.

\begin{figure}
\centering \subfigure[Graph $G \equiv (V,
E)$]{\includegraphics[height=1.33in,trim=1.7in 2.2in 2.6in
1.5in,clip=true]{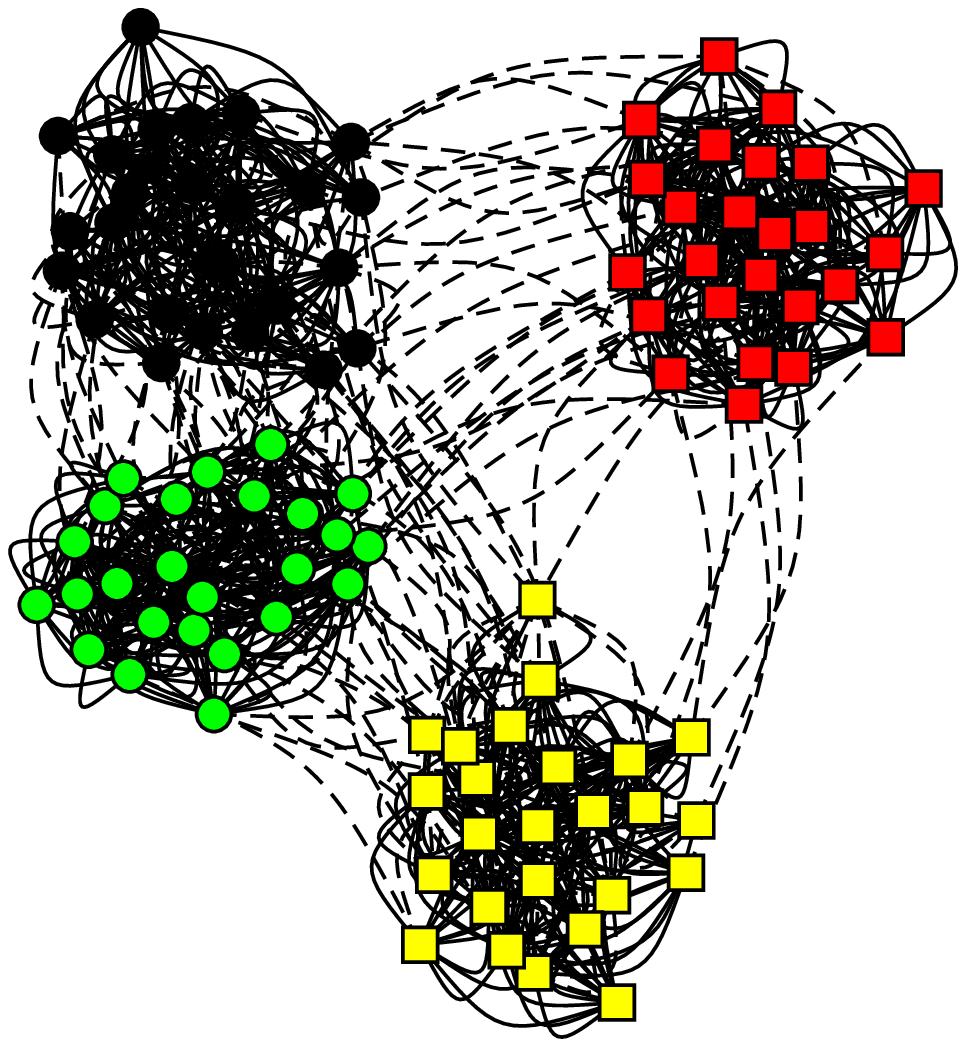} \label{fig:graph_to_distrA}}
\subfigure[$b_L(e)$ distribution, $e \in
E$]{\includegraphics[height=1.33in]{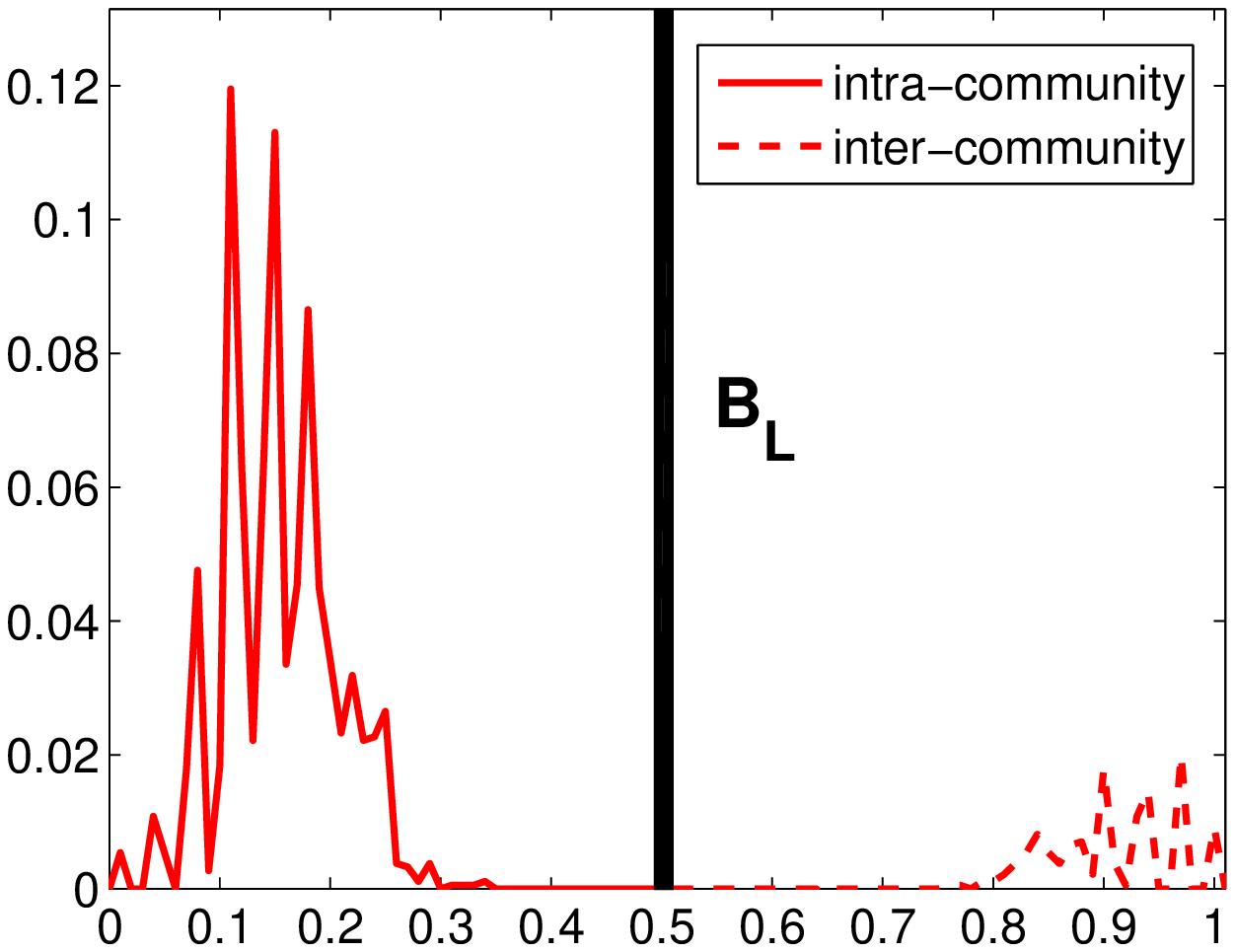}
\label{fig:graph_to_distrB}} \caption{Relation of edge position in
the graph and local bridging $b_L$ probability distribution function
(pdf). Edges drawn with dashed lines on the network of Figure
\ref{fig:graph_to_distrA} are also the ones with the highest local
bridging values (the part of the distribution in Figure
\ref{fig:graph_to_distrB} plotted in dashed line).}
\label{fig:graph_to_distr}
\end{figure}

In order to derive a decision threshold $B_L$ for identifying the
bridge edges of the graph (see line \ref{alg:condition} of Algorithm
\ref{alg:local_bb}), one needs to inspect the distribution of $b_L$
values among the edges of the graph. Figure \ref{fig:graph_to_distr}
illustrates how the position of edges on a graph with community
structure affects their local bridging values. The graph of Figure
\ref{fig:graph_to_distrA} was generated to comprise a synthetic
four-community structure. Edges that link different communities with
each other, i.e. \emph{inter-community} edges, are drawn in dashed
line. According to the distribution of Figure
\ref{fig:graph_to_distrB}, these edges are characterized by high
$b_L$ values, therefore they can be separated by means of
thresholding from the intra-community edges.

The exact probability distribution function of $b_L$ for a given
graph is available only after computing the local bridging function
for each edge of the graph, introducing in that way a global graph
processing step in the Bridge Bounding methodology. However, this
step does not impose severe restrictions on the computational
process. First, according to Equation \ref{eq:local_bridging}, the
computation of $b_L$ can be carried out in a streaming fashion,
since only the neighborhoods of the two endpoints of each edge are
required during the computation. To further reduce the computational
requirements, it is possible to derive an approximation of the $b_L$
probability distribution by computing the local bridging values of a
small random subset of the network edges. Finally, one could even
completely skip the distribution estimation step if it has been
already performed for a graph of similar nature in the past (in
which case one could reuse the previously estimated threshold
$B_L$).

\begin{figure}
\centering \subfigure[$b_L$
distribution]{\includegraphics[width=1.56in]{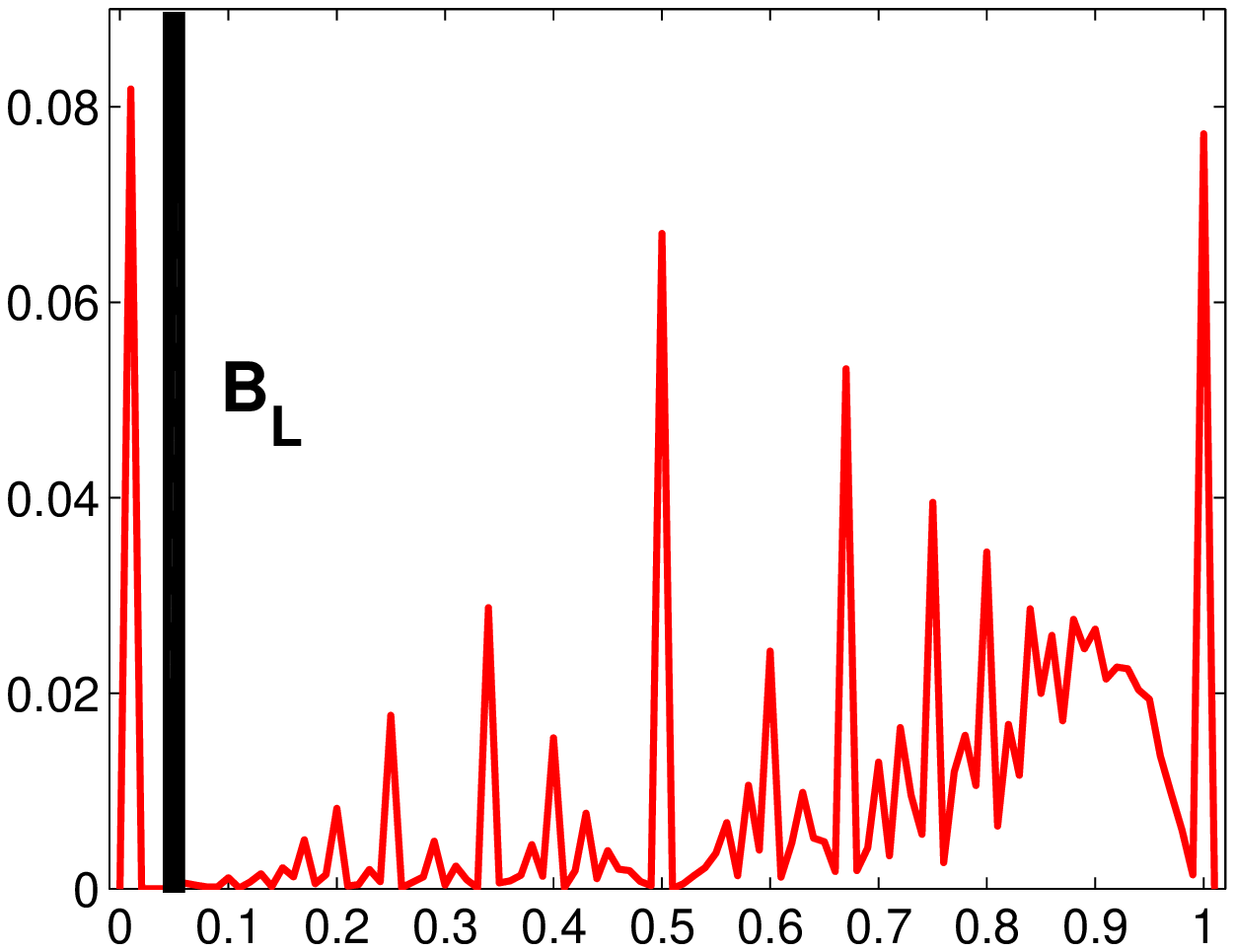}
\label{fig:ecc_distr}} \subfigure[$b_L'$ distribution,
$\alpha=0.7$]{\includegraphics[width=1.56in]{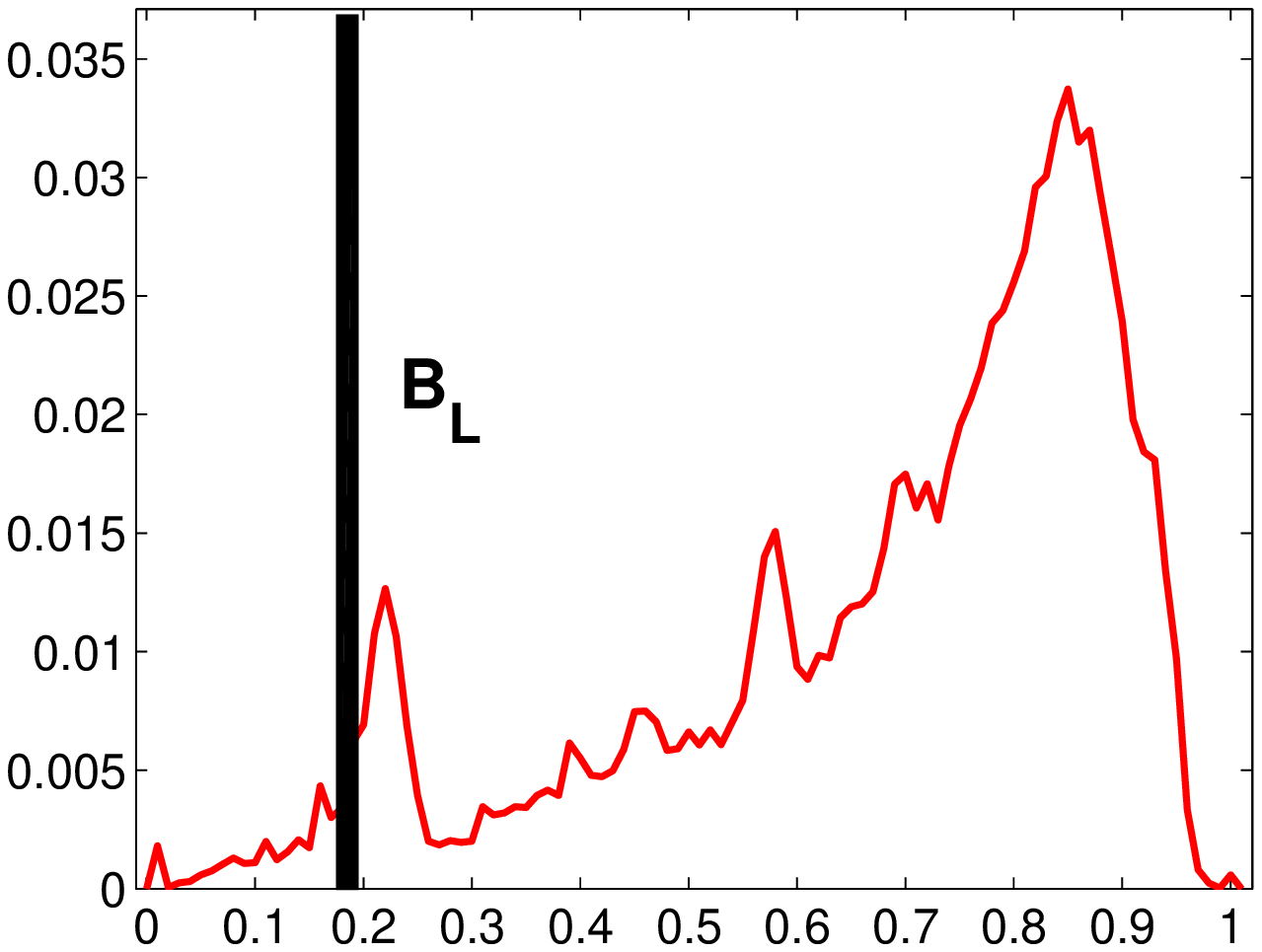}
\label{fig:ecc2_distr}} \caption{Distributions of first-order
($b_L$) and second-order ($b_L'$) local bridging on the English tag
network of Section \ref{sec:lycosIQ}. Note that due to the $b_L$
distribution shape, it is impossible to select a value for $B_L$
such that less than $8\%$ of the network edges are considered
intra-community.} \label{fig:ecc_vs_ecc2}
\end{figure}

\begin{figure}
\centering
\includegraphics[width=3.2in,trim=0in 2.3in 1.2in 1in,clip=true]{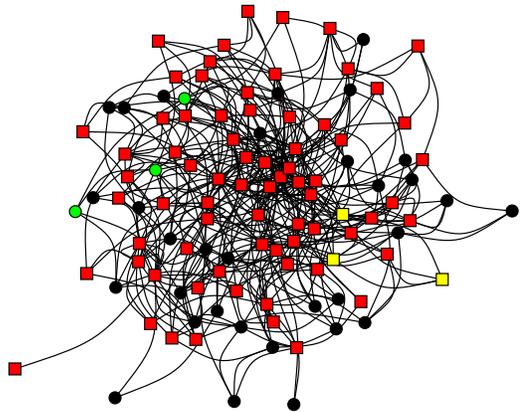}
\caption{Community structure found by Bridge Bounding on a
$100$-node scale-free graph. The structure consists of one large
(red squares), two small (green circles, yellow squares) and $32$
singleton communities (black circles).}
\label{fig:scaleFreePartition}
\end{figure}

The simple measure of local bridging (Equation
\ref{eq:local_bridging}) employed by Bridge Bounding is ideal for
networks with very clear community structure (such as the one of
Figure \ref{fig:graph_to_distrA}). However, the measure is often not
well-suited for detecting communities in real networks. In
particular, when a network is characterized by scale-free topology,
the distribution of $b_L$ values will have a \emph{spiky} shape,
similar to the one in Figure \ref{fig:ecc_distr}, where the depicted
$b_L$ distribution comes from the English LYCOS iQ tag network of
Section \ref{sec:lycosIQ}. In such cases, it is hard to
differentiate between bridge and non-bridge edges. For instance,
according to Figure \ref{fig:ecc_distr}, $8\%$ of the network edges
have local bridging $b_L=0$, thus $\forall B_L \geq 0$, Algorithm
\ref{alg:local_bb} will always consider $8\%$ of the network edges
as non-bridges. In networks with scale-free topology (which commonly
emerge in practice), such a decision would cause Bridge Bounding to
detect a community structure that consists of one large community
and many \emph{singleton} communities, i.e. communities comprising
just one member. The reason for such an outcome is that scale-free
networks maintain a large connected component even when a large
fraction of their edges are removed\footnote{Although Bridge
Bounding does not explicitly remove edges from the underlying
network, it treats bridging edges as bounds, i.e. as non-existent.}
\cite{www:albert00}. Figure \ref{fig:scaleFreePartition} illustrates
the output of Bridge Bounding on a scale-free graph generated by the
\emph{preferential attachment} model of Barab\'{a}si-Albert
\cite{www:barabasi99}.

In order to alleviate this problem, we consider the $2^{nd}$ order
local bridging of an edge $e$, $b_L'(e)$, by computing the weighted
sum (with a mixing parameter $\alpha$) of its local bridging,
$b_L(e)$ and the mean local bridging of the edges constituting its
neighborhood:

\begin{equation}
b_L'(e_{st})=\alpha \cdot
b_L(e_{st})+(1-\alpha)\frac{1}{|N(e_{st})|}\sum_{e \in
N(e_{st})}b_L(e) \label{eq:local_bridging_2}
\end{equation}

By applying Equation \ref{eq:local_bridging_2}, we carry out a
smoothing of the local bridging function by taking into account the
values of the function in the neighborhood around a given edge. The
$\alpha$ parameter defines the extent to which the values of the
neighboring edges are taken into account in the computation of
$b_L'$. Figure \ref{fig:ecc2_distr} illustrates the distribution of
$b_L'$ (using $\alpha=0.7$) for the LYCOS iQ English tag network of
Section \ref{sec:lycosIQ}. Low $b_L'$ values are distributed more
evenly in comparison to the $b_L$ ones. Hence, it is possible to
select a value for $B_L$ such that only a very-low fraction of edges
are considered as intra-community ($\simeq 1\%$ in this example).

Effectively, the computation of $2^{nd}$ order local bridging makes
use of topological information from a wider neighborhood around a
given edge in comparison to local bridging. Following this, one
could consider the $\nu^{th}$ order local bridging, $b_L^{(\nu)}$,
which for sufficiently high values of $\nu$, would utilize
topological information from the whole graph. Obviously, since the
computation of $b_L^{(\nu)}$ is carried out in an iterative manner,
the complexity of computing the measure increases with its order
$\nu$.

In terms of complexity, a progressive global community detection
scheme based on Bridge Bounding is decomposed in two steps: (a)
local network topology function computation and (b) community
detection. Computing the basic local bridging measure for a graph of
$n$ nodes and $m$ edges with average node degree $\overline{d}$ has
a complexity of $O(\overline{d}^2 \cdot m)$ since for each edge, we
need to find the intersection of two sets of average size
$\overline{d}$.\footnote{For the computation of higher-order local
bridging, the complexity raises to $O(\nu \cdot \overline{d}^2 \cdot
m)$. However, we consider that most applications of Bridge Bounding
will make use of second- or at most third-order local bridging
functions.} The community detection step has a complexity of
$O(\overline{d} \cdot n)$, when Algorithm \ref{alg:local_bb} is used
in the global community detection scheme described in
\ref{sec:basic_notation}, since for each node of the network
$\overline{d}$ candidate nodes are considered as candidates for
assignment to the community that is currently being created. Thus,
in total, Bridge Bounding scales with $O(\overline{d}^2 \cdot m+
\overline{d} \cdot n)$.

\section{Evaluation}
\label{sec:evaluation}

In this section, we present a series of experiments we carried out
in order to gain insights into the performance of the proposed
approach. The first part of the experiments compares the performance
of progressive global community detection based on Bridge Bounding
with the one achieved by the community detection method of
Girvan-Newman \cite{www:girvan02}. This comparison is carried out on
synthetic networks with known (predefined) community structure, thus
giving the possibility for objective measurement of the method
performance. In the second part of the experiments, we aim at
gaining insights into real-world complex networks. Therefore, we
apply our community detection technique on two networks created from
the user tagging activities in the English and German version of the
LYCOS iQ question / answering application. Since there is no ground
truth concerning the community structure of the LYCOS iQ tag
network, we use our subjective judgement in order to draw
conclusions on the performance of the proposed method.

\subsection{Synthetic networks}
\begin{figure}
\centering \subfigure[$p_{out}=0.01$]{
\includegraphics[width=1.5in,trim=1in 2.5in 3.2in 2.2in,clip=true]{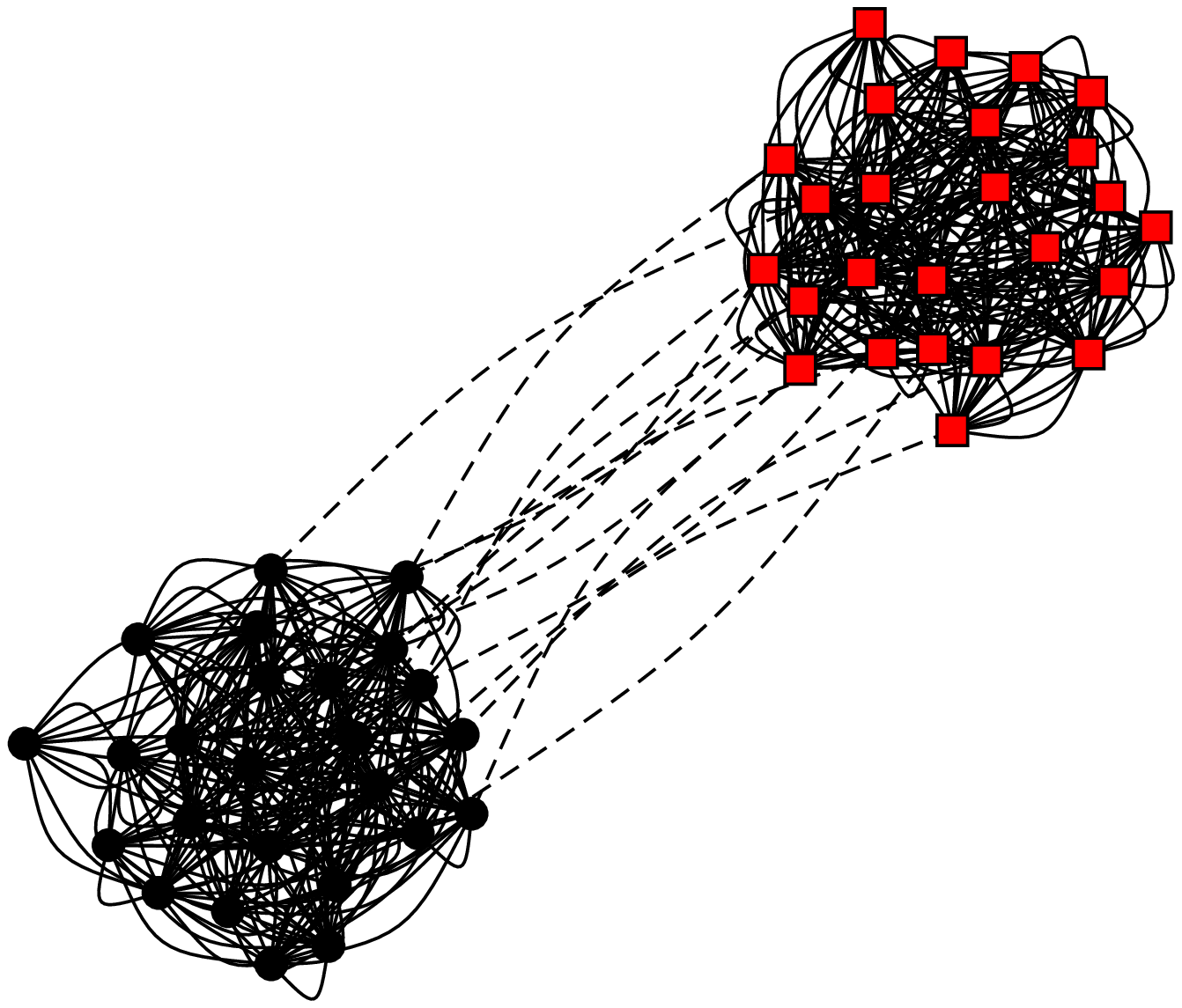} \label{fig:ex_com1}}
\subfigure[$p_{out}=0.08$]{
\includegraphics[width=1.5in,trim=2in 3.5in 3in 1.5in,clip=true]{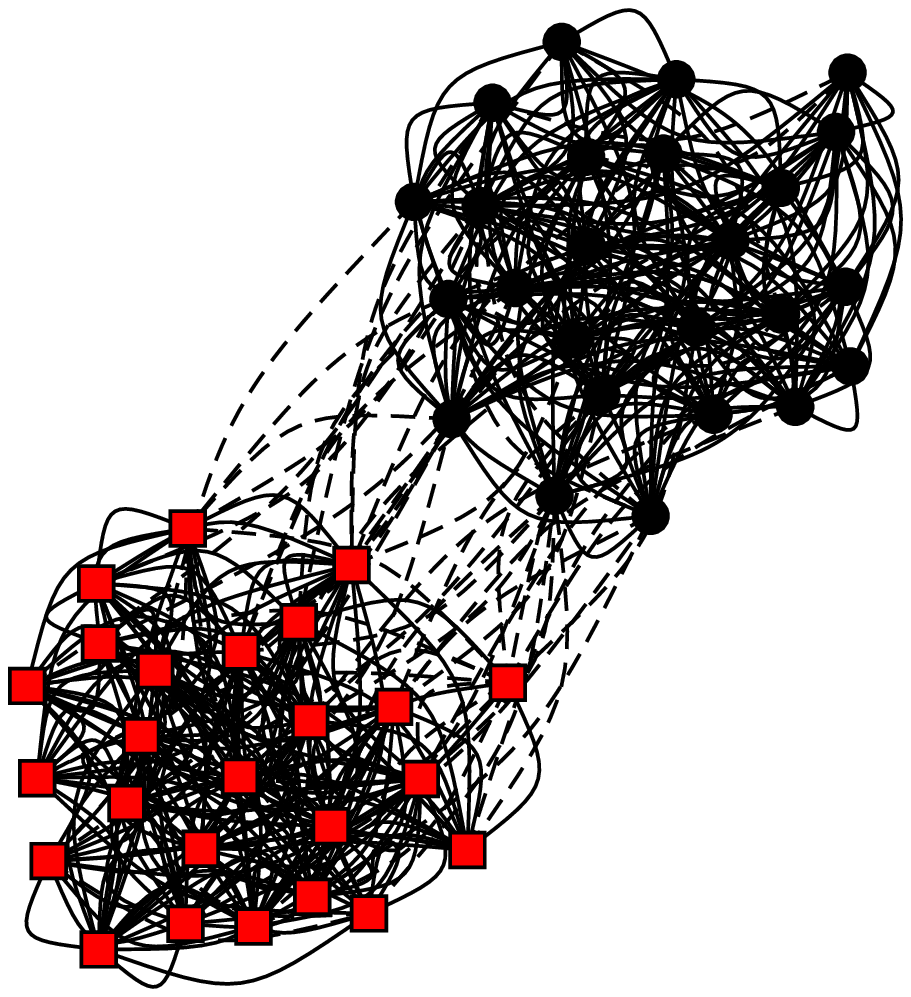} \label{fig:ex_com2}}
\caption{Sample synthetic mixtures of communities generated using
the same set of parameters $\{N=50$, $K=2$, $z_{tot}=18\}$ but
different values for $p_{out}$.} \label{fig:synthetic_communities}
\end{figure}

We created a parameterized community mixture generator following the
strategy described in \cite{www:newman04a} and \cite{www:danon05b}.
According to this, the generation process results in a network with
$N$ nodes which consists of $K$ communities. We control the average
degree $z_{tot}$ of the network nodes, as well as the probability
$p_{out}$ that a node's edge will connect to a node of a different
community. Thus, out of the $z_{tot}$ edges of each node (on
average), $z_{out}=p_{out} \cdot z_{tot}$ edges connect the node to
nodes of different communities. Obviously, higher values of
$p_{out}$ will lead to networks with less profound community
structure. Figure \ref{fig:synthetic_communities} depicts the
difference in the conspicuity of community structure in relation to
the fraction $p_{out}$ of inter-community edges. This network
generation process can be described by a four-element parameter set
comprising $N$, $K$, $z_{tot}$, and $p_{out}$. We also consider a
fifth parameter, namely the \emph{community size variation}
$s_{var}$, which is calculated as the ratio of the biggest community
size to the size of the smallest one. In this case, each community
$C_i$ will have a different average node degree $z_{tot}^i$ and
therefore we define $z_{tot}=\frac{1}{K} \cdot \sum_{i = 1}^K
{z_{tot}^i}$. In the end, we consider the five-element parameter
set:

\begin{equation}
\label{eq:parameter_set} S_{PAR}=\{N, K, z_{tot}, p_{out}, s_{var}\}
\end{equation}

Two widely used measures to evaluate the effectiveness of data
partitioning methods, e.g. community detection, when the \emph{true}
partition structure is known (which is the case when testing with
synthetic networks) are (a) the fraction $F_C$ of correctly
classified instances \cite{www:newman04a} and (b) the
\emph{Normalized Mutual Information} ($\text{NMI}$) introduced in
\cite{www:fred03} and applied for the evaluation of community
detection in \cite{www:danon05b}. Consider two partitions of the
$n$-node graph, $\textbf{P}^a = \{C_0^a, C_1^a, ..., C_{K_a}^a\}$
(true community structure) and $\textbf{P}^b = \{C_0^b, C_1^b, ...,
C_{K_b}^b\}$ (community structure found by algorithm). The fraction
$F_C$ of correctly classified instances is straightforward to
compute only when $K_a=K_b=K$. When the true number of communities
$K_a$ differs from the number of communities $K_b$ found by the
algorithm, we need to first identify a subset of the found
communities $\textbf{P}_c^b \subseteq \textbf{P}^b$, that can be
matched to a subset of the true communities, $\textbf{P}_c^a
\subseteq \textbf{P}^a$. We consider two communities as matching if
they present overlap of more than $50\%$. Then, assuming that
community $C_x^a \in \textbf{P}_c^a$ is the matching community of
$C_i^b \in \textbf{P}_c^b$, $F_C$ is computed by the following
equation.

\begin{equation}
F_C=\frac{1}{n} \cdot \sum_{C_i^b \in \textbf{P}_c^b} {|C_x^a
\bigcap C_i^b|}
\end{equation}

The Normalized Mutual Information between the true partition,
$\textbf{P}^a$, and the one found by the algorithm, $\textbf{P}^b$,
quantifies the extent to which they are similar to each other from
an information-theoretic point of view \cite{www:fred03}.

\begin{equation} \label{eq:nmi}
\text{NMI}(P^a, P^b)=\frac{-2 \cdot \sum_{i=1}^{K_a}
\sum_{j=1}^{K_b} n_{ij}^{ab} log(\frac{n_{ij}^{ab} \cdot n}{n_i^a
\cdot n_j^b})}{\sum_{i=1}^{K_a}n_i^a log(\frac{n_i^a}{n}) +
\sum_{j=1}^{K_b} n_j^b log(\frac{n_j^b}{n})}
\end{equation}

In Equation \ref{eq:nmi}, $n_i^a$ and $n_j^b$ denote the number of
nodes in communities $C_i^a$ and $C_j^b$ respectively, and
$n_{ij}^{ab}$ denotes the number of shared nodes between communities
$C_i^a \in \textbf{P}^a$ and $C_j^b \in \textbf{P}^b$. In general,
$\text{NMI}$ is preferred to the simplistic $F_C$ measure, since it
handles gracefully the cases where $K_a \neq K_b$. $F_C$ is
presented here together with $\text{NMI}$ mainly due to the ease in
its interpretation.

To demonstrate the effectiveness of Bridge Bounding in detecting the
underlying community structure of networks, we compare the
performance of the progressive global community detection scheme
(see Section \ref{sec:basic_notation}) based on Bridge Bounding in
terms of both $F_C$ and $\text{NMI}$ to the performance of the
community detection method by Girvan and Newman (GN)
\cite{www:girvan02} on a multitude of synthetic networks. Since the
GN method employs a divisive approach, it results in a hierarchical
community structure, which contains multiple graph partitions to
communities. Therefore, we needed to select a single partition from
the hierarchy, which we would use to evaluate the performance of the
method. The strategy used by Newman and Girvan in
\cite{www:newman04a} to make this selection is to calculate the
\emph{modularity} $Q$ of each partition and select the partition
which maximizes it.

The modularity of a network partition into $K$ communities is
calculated from the $K \times K$ symmetric matrix $\textbf{e}$ whose
element $e_{ij}$ is the fraction of all edges in the network that
link vertices in community $i$ to vertices in community $j$.
Further, we define the row (or column) sums $\alpha_i=\sum_j e_{ij}$
which represent the fraction of edges that connect to vertices in
community $i$. Based on the above definitions, the measure of
modularity is defined as:

\begin{equation}
\label{eq:modularity} Q = \sum_i (e_{ii}-\alpha_i^2)
\end{equation}

This quantity measures the fraction of edges in the network that
connect vertices of the same community (i.e. intra-community edges)
minus the expected value of the same quantity in a network with the
same community partition but random connections between the
vertices. If the number of intra-community edges is no better than
random, we would get $Q=0$. For perfect separation to communities
(i.e. communities that are completely disconnected from each other
on the graph), we get $Q=1$. In practice, modularity values in the
range from $0.3$ to $0.7$ indicate significant community structure.

\begin{table}
\centering \caption{Comparison of performance between a global
scheme based on Bridge Bounding with local bridging (BB), Bridge
Bounding with $2^{nd}$ order local bridging (BB') and the method of
Girvan and Newman (GN) \protect \cite{www:girvan02}. The performance
is measured on synthetic networks generated using the set $S_{PAR}^1
= \{200, 4, 40, p_{out}, 1.0\}$ of parameters, with $p_{out}$ being
the free parameter.}
\begin{tabular}{|l|c|c|c|c|c|c|} \hline
 & \multicolumn{3}{|c|}{$F_C$} & \multicolumn{3}{|c|}{$\text{NMI}$} \\ \hline
$p_{out}$ & BB & BB' & GN & BB & BB' & GN \\ \hline 0.01 & 100 & 100 & 100 & 1.0 & 1.0 & 1.0 \\
\hline 0.05 & 100 & 100 & 100 & 1.0 & 1.0 & 1.0 \\ \hline 0.1 & 100
& 100 & 50 & 1.0 & 1.0 & 0.86 \\ \hline 0.15 & 100 & 99 & 50 & 1.0 &
.98 & 0.86 \\ \hline 0.20 & 99 & 74 & 50 & 0.98 & 0.84 & 0.86
\\ \hline 0.25 & 24 & 24 & 0 & 0.54 & 0.56 & 0.02
\\ \hline
\end{tabular} \label{tab:perf_1}
\end{table}

\begin{table}
\centering \caption{Similar comparison of performance as in Table
\ref{tab:perf_1}, but on synthetic networks that were generated
using the set $S_{PAR}^2 = \{200, 4, 40, 0.01, s_{var}\}$ of
parameters, with $s_{var}$ being the free parameter.}
\begin{tabular}{|l|c|c|c|c|c|c|} \hline
 & \multicolumn{3}{|c|}{$F_C$} & \multicolumn{3}{|c|}{$\text{NMI}$} \\ \hline
$s_{var}$ & BB & BB' & GN & BB & BB' & GN \\ \hline 1.1 & 100 & 100 & 100 & 1.0 & 1.0 & 1.0 \\
\hline 1.5 & 100 & 100 & 100 & 1.0 & 1.0 & 1.0 \\ \hline 1.6 & 99.5
& 100 & 100 & 0.99 & 1.0 & 1.0 \\ \hline 1.7 & 88 & 98 & 100 & 0.82
& 0.96 & 1.0 \\ \hline 1.8 & 85.5 & 97 & 100 & 0.79 & 0.95 & 1.0
\\ \hline 1.9 & 58.5 & 87 & 90 & 0.68 & 0.82 & 0.88 \\
\hline 2.0 & 12.5 & 80 & 82 & 0.45 & 0.73 & 0.81 \\
\hline 2.5 & 0 & 62 & 75 & 0.45 & 0.63 & 0.72 \\ \hline
\end{tabular} \label{tab:perf_2}
\end{table}

We created two sets of networks containing synthetic communities.
The first set of such networks was generated holding the four
network generation parameters of Equation \ref{eq:parameter_set}
constant and varying $p_{out}$. This is a widely adopted test
process \cite{www:newman04a, www:radicchi04, www:danon05b} to test
the performance of a community detection method as the communities
of the synthetic graph gradually become less well-separated. Table
\ref{tab:perf_1} presents the comparison between the performance of
Bridge Bounding (by use of both first- and second-order local
bridging) and the GN method \cite{www:girvan02}. Both Bridge
Bounding methods present equally good or better performance than GN
across the range of $p_{out}$ values that were used for testing.

A further test involved the generation of an additional set of
networks by varying the $s_{var}$ parameter in order to end up with
networks comprising communities of unequal sizes. Table
\ref{tab:perf_2} provides an overview of the results obtained from
the three methods of our study. Apparently, the use of the local
bridging function (Equation \ref{eq:local_bridging}) becomes
problematic for Bridge Bounding as soon as the size variation among
the underlying communities exceeds a certain value (e.g. for
$s_{var} \geq 2$, we measured $\text{NMI}(BB) < 0.5$). In contrast,
Bridge Bounding with the use of $2^{nd}$ order local bridging as
well as the GN method yielded consistently better results in this
series of tests. Hence, it becomes clear that the use of more
sophisticated local topology measures, such as the $2^{nd}$ order
local bridging, could be crucial for the success of the proposed
method.

\subsection{LYCOS iQ tag network}
\label{sec:lycosIQ} LYCOS iQ is a collaborative question/answering
application where people ask and answer questions on any topic. The
application is available in six languages, German, English, French,
Danish, Swedish and Dutch with German attracting the largest
community of users. In order to support the users' efforts of
searching for relevant questions, the application incorporates a
tagging functionality, similar to the one used in typical
\emph{social tagging systems} such as
delicious\footnote{http://delicious.com} and
flickr\footnote{http://flickr.com}. There are no static categories
and tags are not predefined by the system, but the users' inputs are
checked against tags existing in the system database to prevent
duplicates.

\begin{comment} Users can add tags to their questions and bookmarks, and they
can select interesting tags to receive automatic recommendations for
thematically connected questions. Further, registered users that
answer questions are related with the questions' tags and can
collect iQ points (internal currency system) for these actions.
Thus, questions, answers, bookmarks and even user profiles can be
browsed and searched by using tags.
\end{comment}

Question submitters have the possibility of attaching more than one
tag to each of their questions. Therefore, it is possible to create
a tag network from the collaborative tagging activities of users. In
this network, the vertex set comprises the tags chosen by users to
tag their questions and the edge set contains the co-occurrences
between tags in the users' questions. When a question is tagged with
more than two tags, then all possible pairwise co-occurrences are
added to the network. For each tag of the network, its frequency
($tf$) is available. Further, the co-occurrence frequency ($cf$)
between each pair of tags is
available. %Hence, the resulting tag network is weighted.

\begin{comment}
\begin{table}
\centering \caption{Tag networks summary. For each network
$G=\{V,E\}$, it is $|V|=tags$ and $|E|=tag\text{-}pairs$.}
\begin{tabular}{|l|c|c|c|} \hline
& $tags$ & $tag\text{-}pairs$ & $questions$\\ \hline English (UK) & 26,758 & 109,340 & 62,497\\
\hline German (DE) & 261,867 & 1,408,989 & 942,405\\ \hline
\end{tabular} \label{tab:tn_summary}
\end{table}
\end{comment}

\begin{figure}
 \centering \subfigure[tag frequency]{
\includegraphics[height=1.16in]{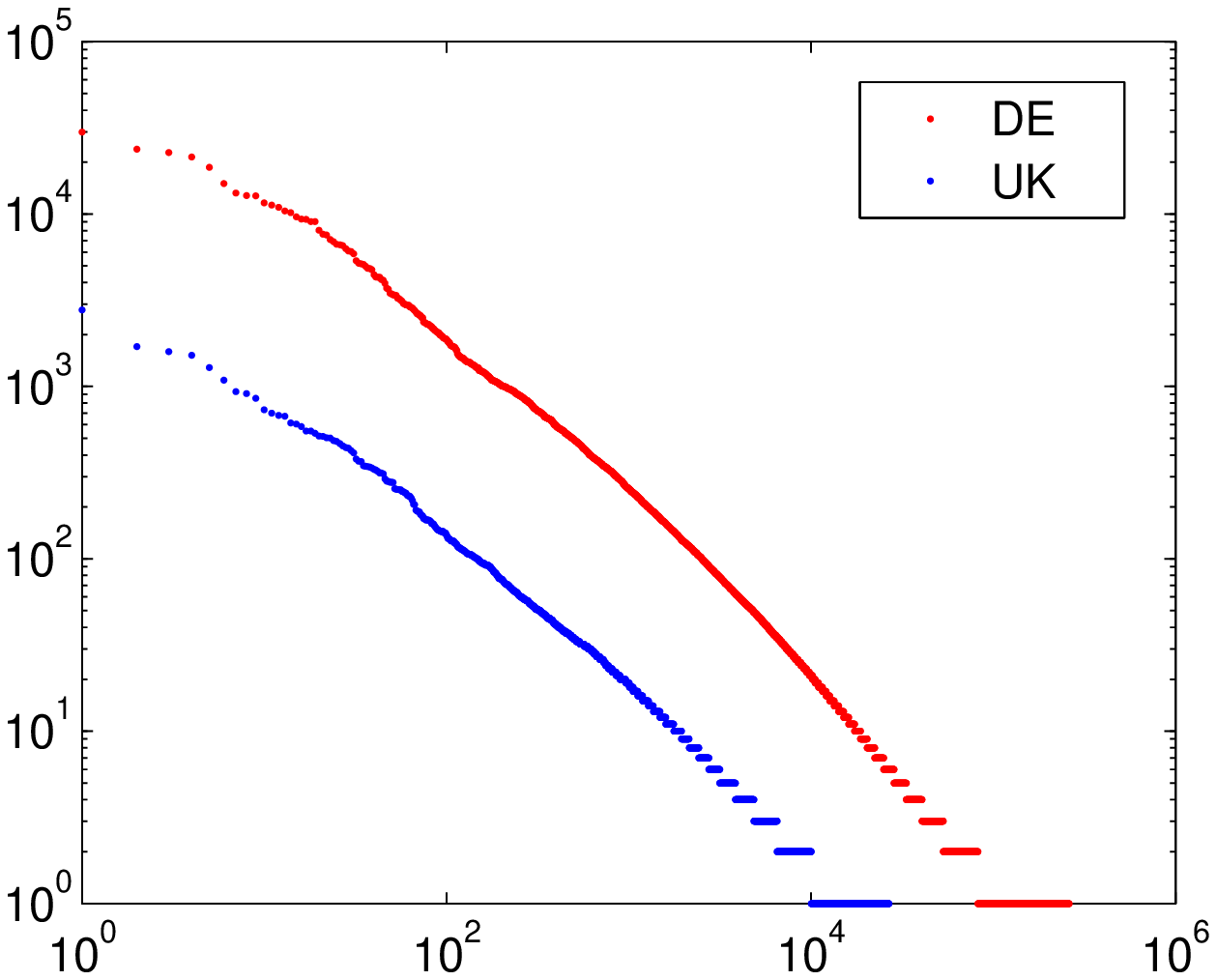} \label{fig:distA}} \subfigure[cooccurrence frequency]{
\includegraphics[height=1.16in]{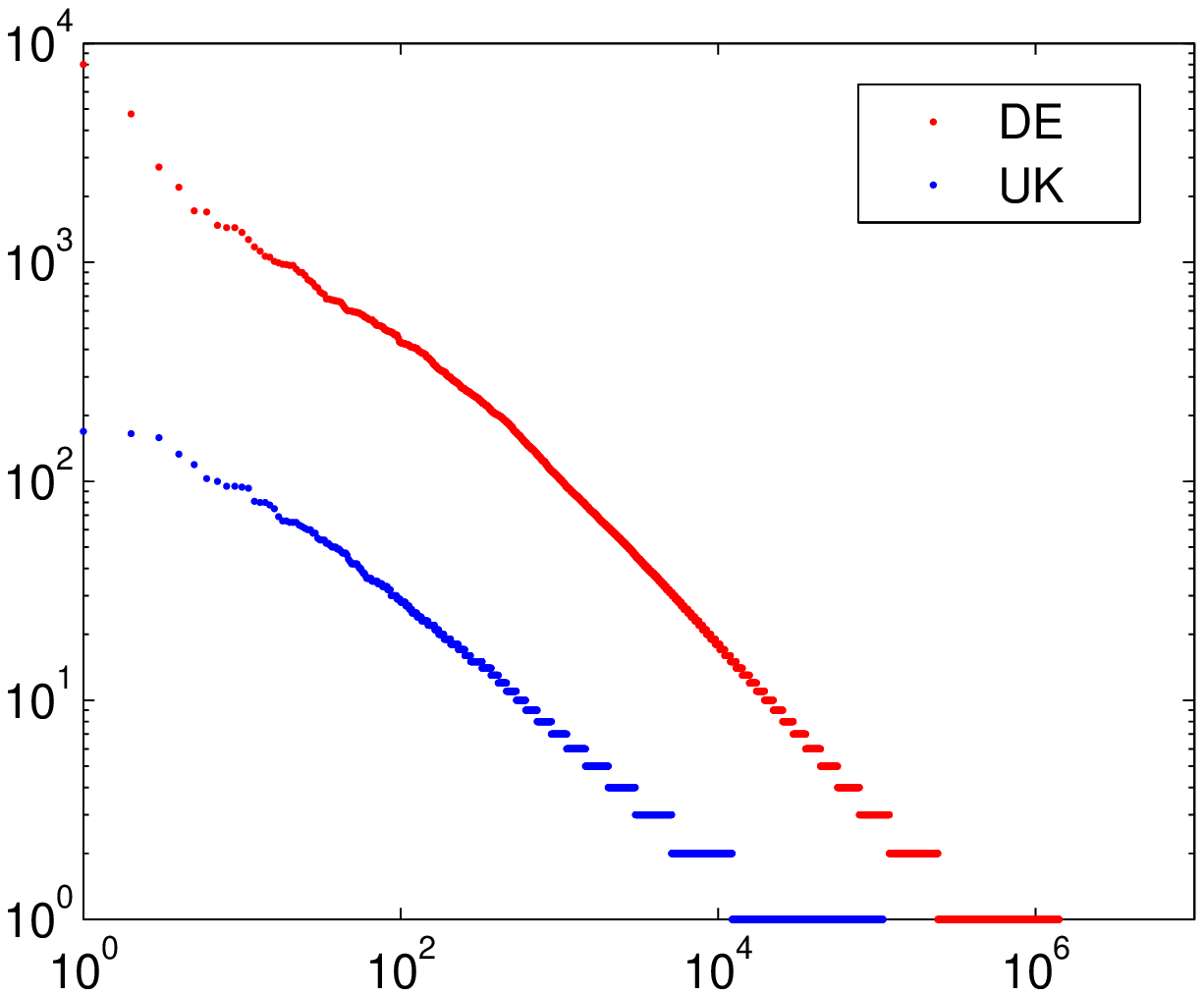} \label{fig:distB}}
\subfigure[cooccurrence frequency]{
\includegraphics[height=1.16in]{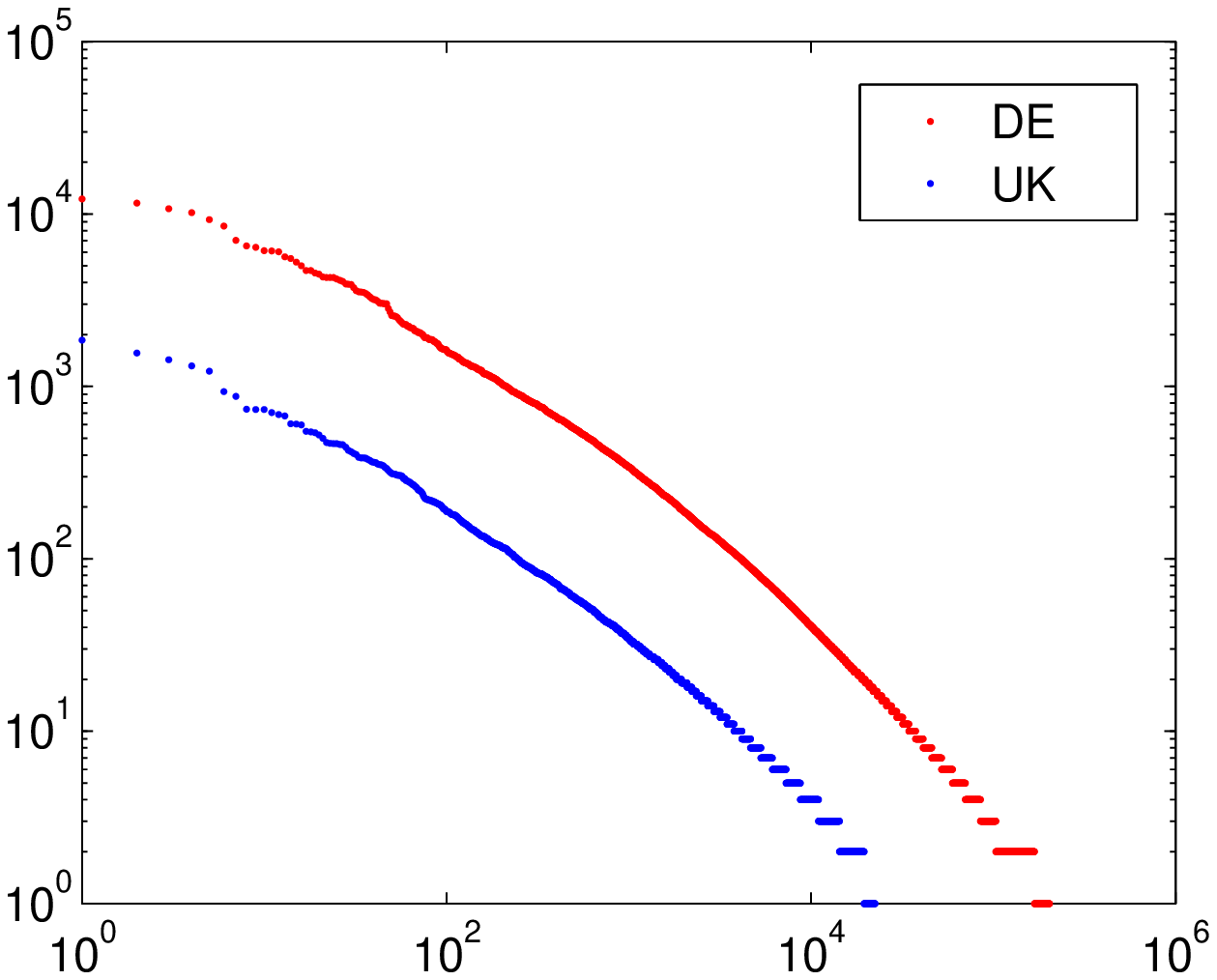} \label{fig:distC}}
\caption{Rank plots of tag, cooccurrence frequencies and node
degrees for the German and English LYCOS iQ tag networks.}
\label{fig:distributions_tf_coc}
\end{figure}

Figure \ref{fig:distributions_tf_coc} illustrates the rank plots of
tag and cooccurrence frequencies as well as of the node degrees
observed in the German and English LYCOS iQ tag networks. A highly
skewed behavior is obvious in the tagging activities of users, e.g.
in the English dataset, a small set of tags is used very frequently
(hundreds of times), while the majority of them is used less than
$10$ times. The frequency of cooccurrence between tags follows a
similar pattern, with less than a thousand tag pairs occurring
together in more than a few questions. Finally, the node degrees
follow a long-tail distribution, indicating that the tag networks
are characterized by scale-free topology. A considerable number of
tags are even disconnected from the rest of the network meaning that
they were used in isolation. To reduce the amount of noisy tags in
the network, we filtered out tags that were either disconnected or
appeared less than twice in the dataset.

\begin{table}
\centering \caption{Tag networks used in this case study. For each
network $G=\{V,E\}$, it is $|V|=tags$ and $|E|=tag\text{-}pairs$.}
\begin{tabular}{|l|c|c|c|} \hline
& $tags$ & $tag\text{-}pairs$ & $questions$\\ \hline English (UK) & 9,517 & 77,243 & 62,497\\
\hline German (DE) & 78,138 & 896,486 & 942,405\\ \hline
\end{tabular} \label{tab:tn_summary}
\end{table}

Table \ref{tab:tn_summary} provides a summary of the two tag
networks that we obtained after the aforementioned filtering step.
The tag network induced from the tagging activity on the German
version of LYCOS iQ, is far larger than the one created from the
English version. Nevertheless, we preferred to present community
snapshots and examples only from the English tag network to ensure
that even readers who are not familiar with German can understand
them. Since the proposed community detection method relies only on
information regarding the network topology, the outcome of the
method is language independent. We could confirm this intuition by
inspecting the community detection results on both tag networks.

\begin{comment}
\begin{table}
\centering \caption{Execution times for community detection in the
English and German tag network. The tests were executed on a Java
Virtual Machine with $512 MB$ of heap memory, running on a Pentium
Xeon $3.2$ GHz.}
\begin{tabular}{|l|c|c|c|} \hline
& $b_L'$ computation (secs) & community detection (secs) & total time (secs) \\ \hline English (UK) & 3.65 & 3.72 & 7.37\\
\hline German (DE) & 124.84 & 365.05 & 489.89\\ \hline
\end{tabular} \label{tab:algorithm_efficiency}
\end{table}
\end{comment}

\begin{figure}
\centering
\includegraphics[width=3in,trim=1.05in 1.2in 0.8in 1.2in, clip=true]{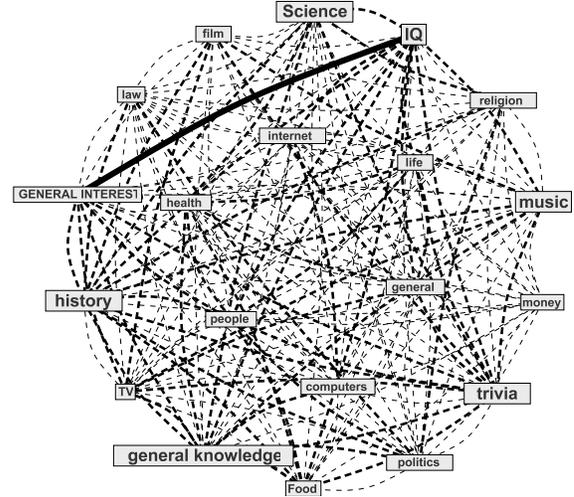}
\caption{Overview of the English LYCOS iQ tag network. We use the
following conventions for tag network visualizations: (a) Font size
is proportional to tag frequency, (b) Edge thickness is proportional
to cooccurrence frequency, (c) Edges identified as bridges are drawn
in dashed line.} \label{fig:tn_overview}
\end{figure}

\begin{figure}
\centering
\includegraphics[width=2.5in]{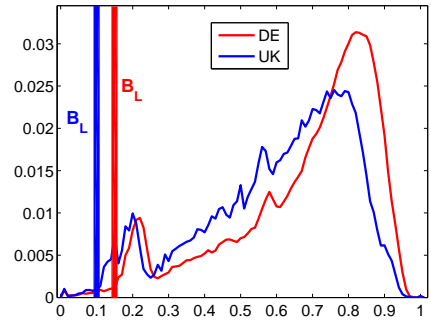}
\caption{Distribution of $b_L'$ values in the two LYCOS iQ tag
networks.} \label{fig:ecc2_tag_net_distr}
\end{figure}

Figure \ref{fig:tn_overview} provides a high-level view of the most
prominent topics coming up through the users' questions in the LYCOS
iQ application. The tags depicted in this view were selected based
on their degree in the network. Although the resulting network is
very densely connected, one can already see that all tags (apart
from the pair ``IQ''-``GENERAL INTEREST'') belong to different
communities (since the dashed edges have been found to be
inter-community edges, after thresholding based on the $b_L'$
distribution of the network in Figure \ref{fig:ecc2_tag_net_distr}).

\begin{figure}
\centering
\includegraphics[width=2.5in,trim=1in 1.1in 0.5in 1.1in,clip=true]{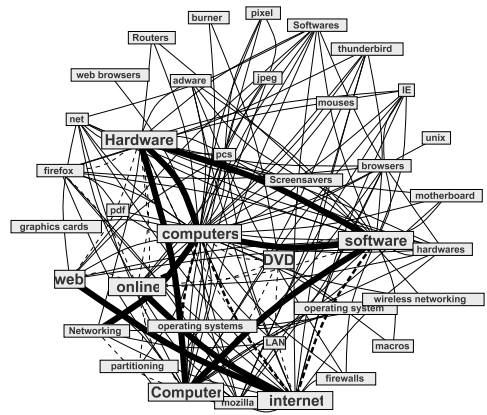}
\caption{Community around tag ``computers''.}
\label{fig:computers_community}
\end{figure}

\begin{figure}
\centering \includegraphics[width=2.5in, trim=0.9in 1.1in 0.5in
1.1in,clip=true]{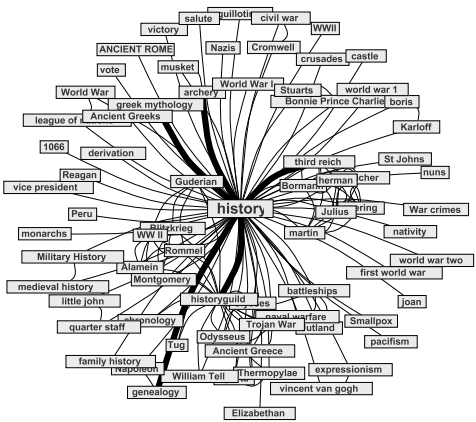} \caption{Community around
tag ``history''.} \label{fig:history_community}
\end{figure}

In order to explore the topic structure of the tag network in more
depth, we selected some of the top-level tags as seed nodes and
inspected the resulting communities. Figures
\ref{fig:computers_community} and \ref{fig:history_community}
present the communities around tags ``computers'' and ``history''.
In both figures it is apparent that most edges are considered
intra-community. Also, note that while the ``computers'' community
is densely connected, the ''history`` community resembles a
star-shaped graph: it remains connected through its central tag
(``history'').

\begin{figure}
\centering \subfigure[Music]{\includegraphics[width=1.56in,trim=0in
1in 1in
0.5in,clip=true]{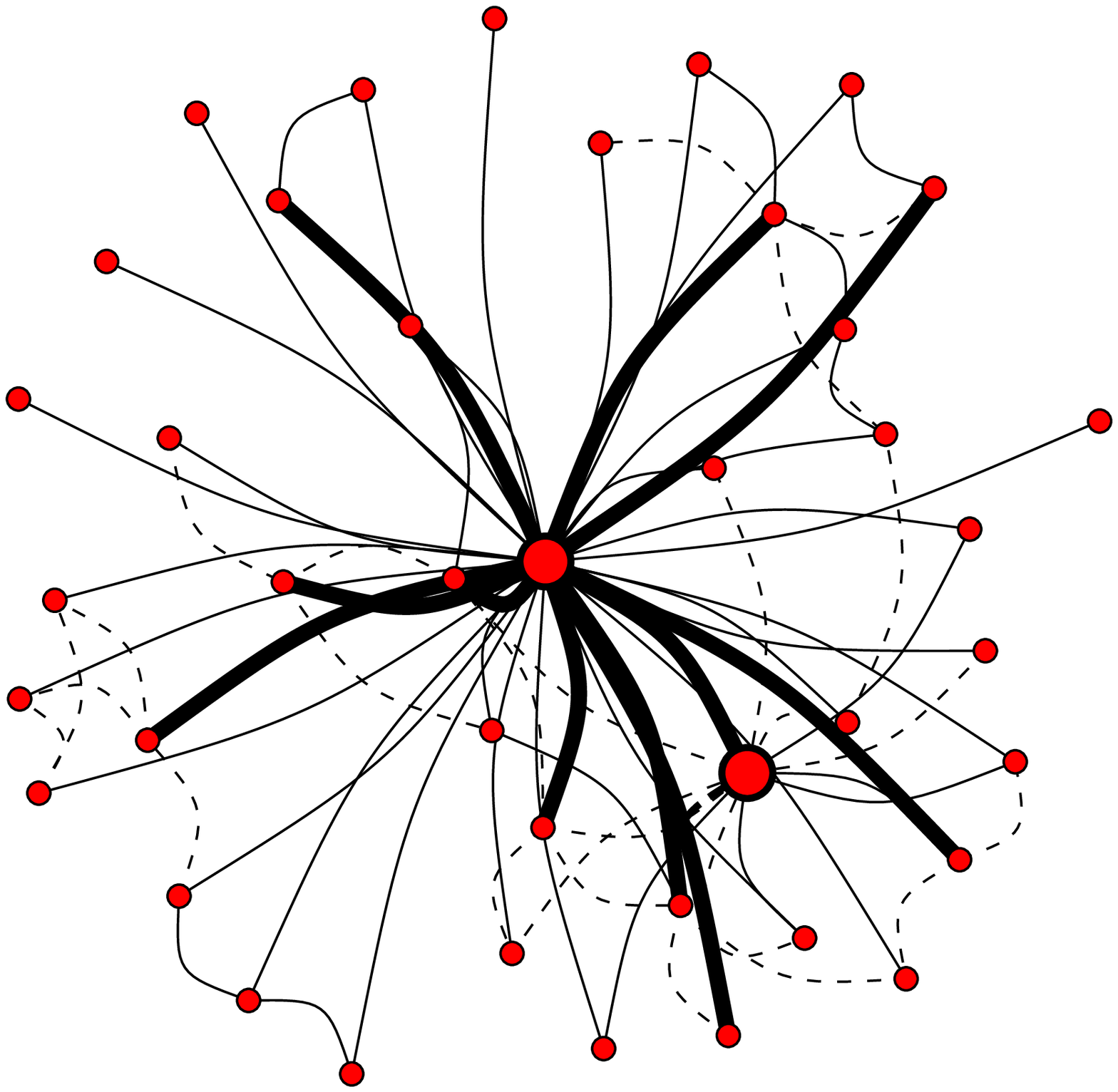}\label{fig:music_community}}
\subfigure[Science]{\includegraphics[width=1.56in,trim=2in 1in 0in
1.5in,clip=true]{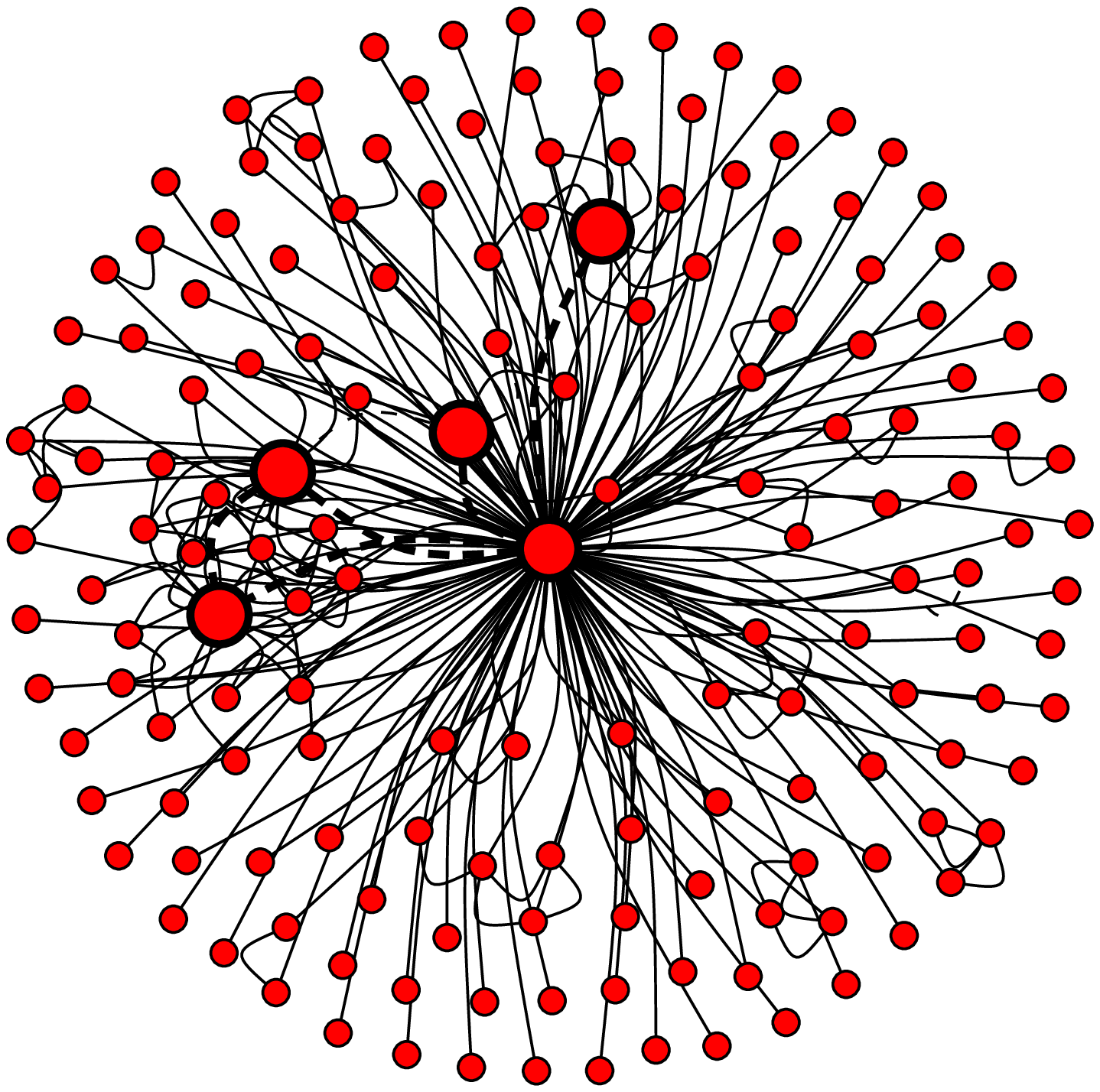}\label{fig:science_community}}
\subfigure[Film]{\includegraphics[width=1.56in,trim=1.7in 0.5in
0.1in 1.2in,
clip=true]{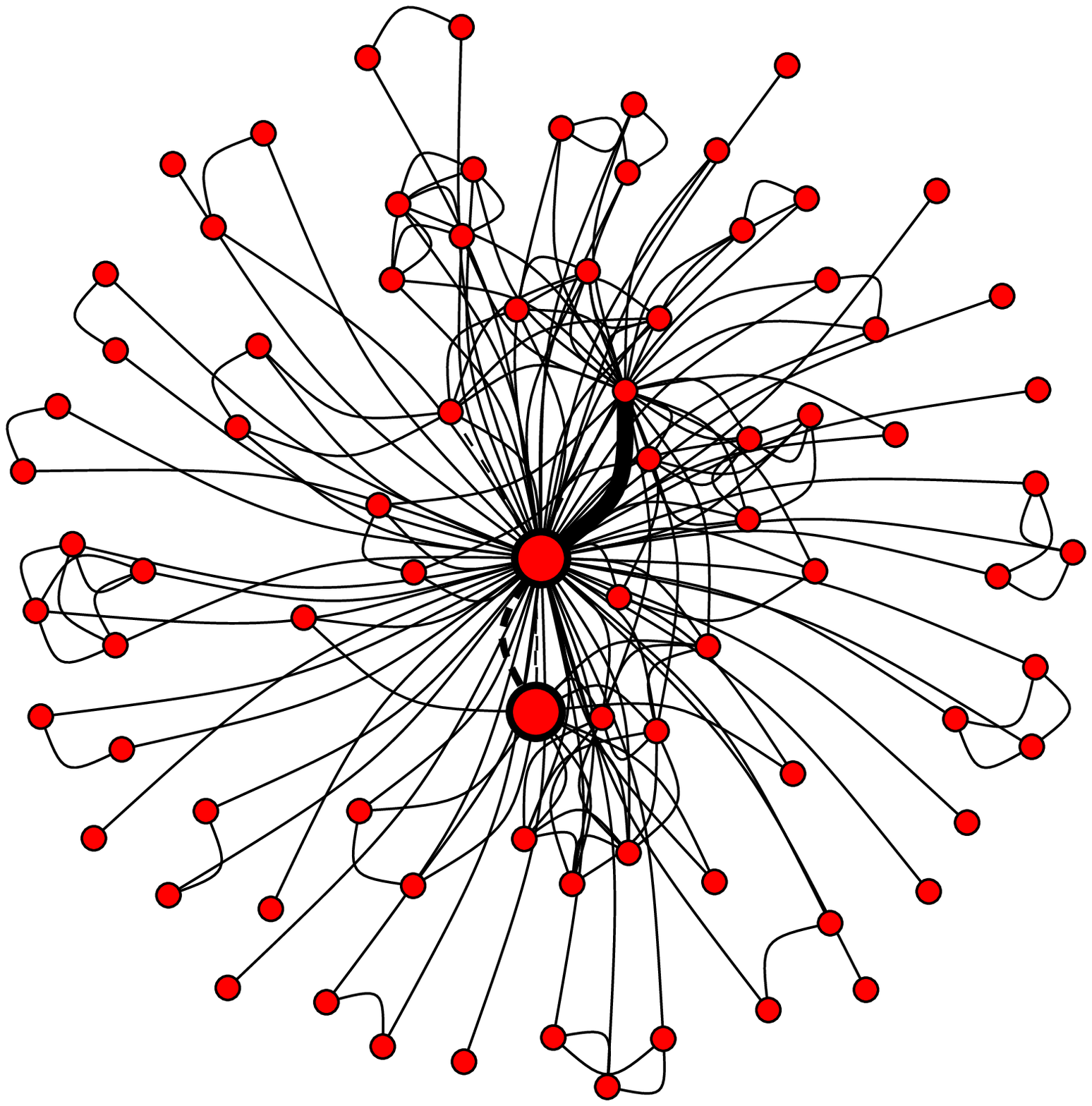}\label{fig:film_community}}
\subfigure[Animals]{\includegraphics[width=1.56in,trim=0.7in 0.8in
0.7in
1.3in,clip=true]{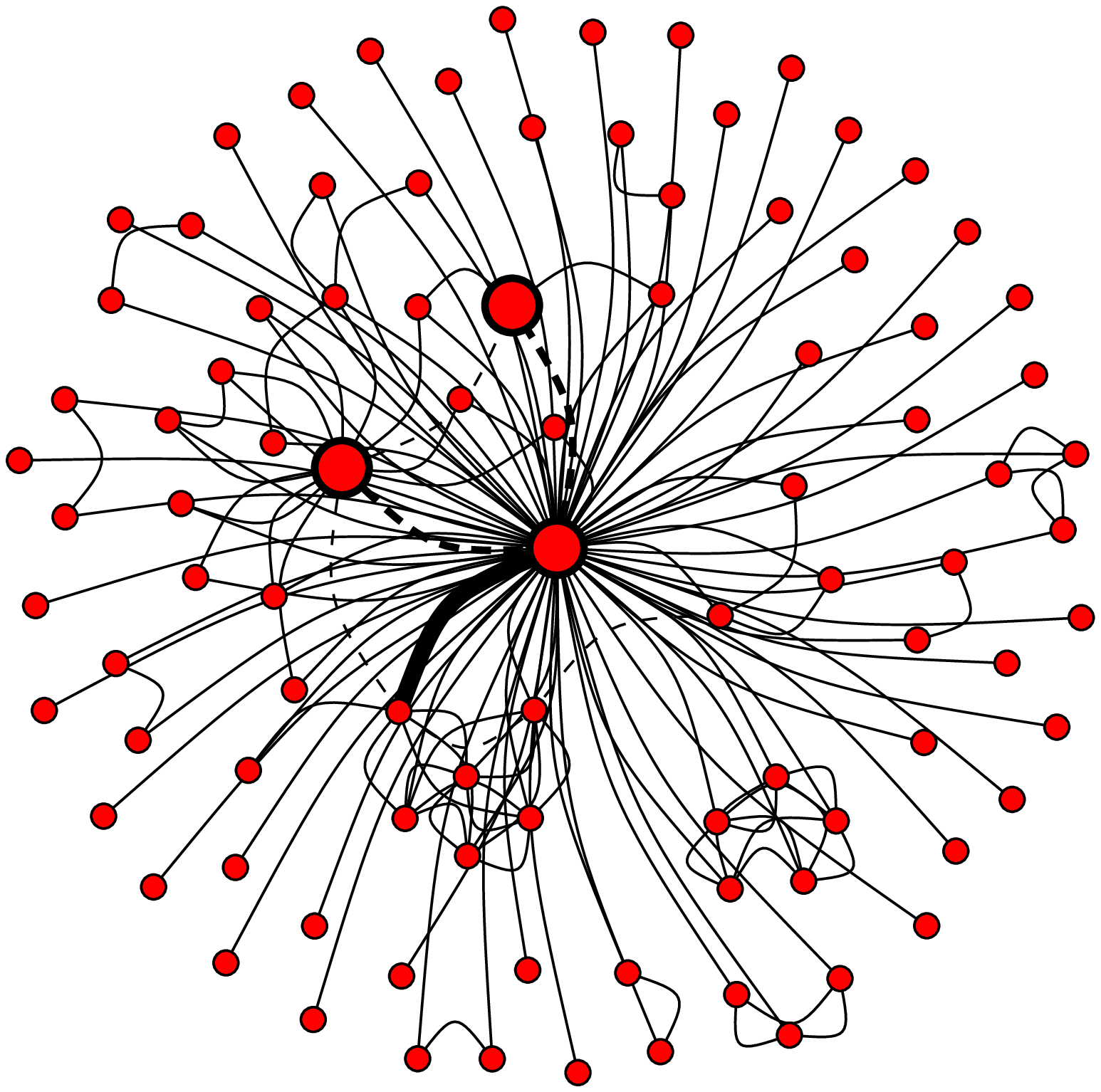}\label{fig:animal_community}}\caption{Further
examples of community shapes. The presented communities were created
using ``music'', ``science'', ``film'' and ``animals'' as seed
nodes.}\label{fig:community_shapes}
\end{figure}

Four additional tag communities are depicted in Figure
\ref{fig:community_shapes}. The complexity of their structure
depends on the topic of the respective community. For instance, the
community formed around the tag ``music'' (Figure
\ref{fig:music_community}) has a much simpler structure than the one
created using ``science'' as the seed tag (Figure
\ref{fig:science_community}). There are two possible reasons for
this: (a) science is a more general topic than music, containing
sub-topics such as physics, medicine, biology and astronomy (these
correspond to the four large nodes of Figure
\ref{fig:science_community}), (b) the questions submitted by LYCOS
iQ users (and consequently the tags used to describe them) are more
focused to particular aspects of music, e.g. pop music artists.

Further, a noteworthy observation regarding the structure of the
communities around ``film'' (Figure \ref{fig:film_community}) and
``animals'' (Figure \ref{fig:animal_community}) is the existence of
small cliques (between 3 and 5 members) within them. Those
correspond to tags related to particular films in the ``film''
community (e.g. ``batman''-``Christian Bale''-``comic'') or tags
related to groups of animals (e.g.
``leopards''-``panthers''-``mammals'') in the ``animals'' community.
This indicates the existence of semantic hierarchies within topics
(e.g. ``mammals'' are a subclass of ``animals''; ``leopards'',
``panthers'' are a subclass of ``mammals''), which could be further
validated by means of machine learning techniques \cite{www:zhou07}.

As stated earlier, detecting the topic communities within a tag
network, similar to the one created from LYCOS iQ application
(nowadays, there are plenty of Web 2.0 applications incorporating
collaborative tagging characteristics), can be beneficial for both
the users and the administrators of the application. Users can be
provided with a community view of the tags that are related to their
context. For instance, when a LYCOS iQ user submits a question to
the system, the text of her question can be parsed and matched
against the tags already available in the system. Then, by
identifying the community (or communities) that her question belongs
to, it is possible to recommend relevant tags for use as descriptors
of the question or relevant questions that have been tagged with
tags belonging to the respective community. Further, administrators
of such applications could use community detection in the context of
a content monitoring and trend tracking framework for supporting the
operation of important administrative tasks, e.g. online ad
targeting or content moderation (which is most frequently synonymous
to spam detection).

\begin{comment}

\begin{figure}
\centering \subfigure[Headache]{\includegraphics[width=1.56in, trim=
1.5in 1.5in 0.9in 1.3in,clip=true]{headache_community.eps}
\label{fig:exTnEng_1}}
\subfigure[NVQ]{\includegraphics[width=1.56in, trim= 0.7in 1.3in
0.4in 1.3in,clip=true]{nvq_community.eps} \label{fig:exTnEng_2}}
\caption{Sample topic communities from the English LYCOS iQ tag
network.} \label{sampleIQnetworksEN}
\end{figure}

\begin{figure}
\centering \subfigure[Bogen]{\includegraphics[width=1.56in, trim=
0.7in 1.3in 0.7in 1.3in,clip=true]{Bogen.eps} \label{fig:exTnDe_1}}
\subfigure[3D software]{\includegraphics[width=1.56in, trim= 0.7in
1.3in 0.5in 1.3in,clip=true]{3D.eps} \label{fig:exTnDe_2}}
\caption{Sample topic communities for the German LYCOS iQ tag
network.} \label{sampleIQnetworksDE}
\end{figure}

\end{comment}

\section{Conclusions}
\label{sec:conclusions} We introduced Bridge Bounding, a local
methodology for community detection in large networks. The
methodology is based on the notion of local network topology
functions to quantify the extent to which edges act as community
boundaries, i.e. bridges. We showed that use of local bridging, a
topology function based on the widely used edge clustering
coefficient, resulted in successful discovery of existing community
structure in synthetic networks, but failed to do so in networks of
scale-free topology. For that reason, we employed the second- and
higher-order local bridging functions to derive smoother estimates
of the bridging properties of edges. The proposed methodology is
extremely efficient, scaling with $O(\overline{d}^2 \cdot m+
\overline{d} \cdot n)$ for networks of $n$ nodes and $m$ edges with
average node degree $\overline{d}$.

A series of tests on synthetic networks with controlled community
structure provides evidence that the Bridge Bounding method (with
use of the $2^{nd}$ order local bridging function) performs equally
well or better than the widely used method of Girvan and Newman.
Moreover, application of our method on two large tag networks coming
from the LYCOS iQ question/answering application proved beneficial
in studying the underlying topic structure and can benefit both
users and administrators of Web 2.0 applications with social tagging
features.

In the future, we plan to carry out more thorough evaluation tests
on the tag communities produced by Bridge Bounding. Specifically, we
plan to conduct a user study among selected LYCOS iQ users in order
to derive manual judgements on the quality of the detected
communities. Subsequently, we are going to consider the potential of
new edge bridging functions and of more sophisticated strategies for
community detection based on Bridge Bounding. Instead of the
currently employed fixed-threshold strategy for deciding whether an
edge is intra- or inter-community, we will test the potential of
adaptive threshold strategies. Finally, we intend to look into
extensions that will endow the method with capabilities for
uncovering hierarchical relations within the community structure.

%ACKNOWLEDGMENTS
\section{Acknowledgments}
This work was supported by the WeKnowIt project, partially funded by
the European Commission, under contract number FP7-215453. We would
also like to acknowledge the use of the
JUNG\footnote{http://jung.sourceforge.net} framework for parts of
our implementation. Finally, we would like to acknowledge the use of
the English and the German tag data sets from the LYCOS iQ
application operated by LYCOS Europe.


\balance
\bibliographystyle{abbrv}
\bibliography{papadop_bridgebounding}  % papadop_bridgebounding.bib is the name of the Bibliography in this case

%\balancecolumns
\balancecolumns % GM July 2000
\end{document}